\documentclass[twocolumn,aps,prb,showpacs,floats,superscriptaddress,amsmath,amssymb]{revtex4}

\usepackage{graphicx}
\usepackage{dcolumn}
\usepackage{bm}

\def\reff#1{(\ref{#1})}

\begin{document}

\title{Stick-slip dynamics of an oscillated sessile drop}

\author{Irina S. Fayzrakhmanova}
\affiliation{Department of Theoretical Physics, Perm State University, Bukirev 15, Perm 614990, Russia}
\affiliation{CFD Laboratory, Institute of Continuous Media Mechanics UB RAS, Korolev 1, 614013 Perm, Russia}
\affiliation{Department of Physics and Astronomy, University of Potsdam, Karl-Liebknecht-Str. 24/25, D-14476 Potsdam-Golm, Germany}

\author{Arthur V. Straube\footnote{Author to whom correspondence should be addressed. Electronic mail: arthur.straube@gmail.com}}
\affiliation{Institut f\"ur Theoretische Physik, Technische Universit\"at Berlin, Hardenbergstr. 36, D-10623 Berlin, Germany}
\affiliation{Department of Physics and Astronomy, University of Potsdam, Karl-Liebknecht-Str. 24/25, D-14476 Potsdam-Golm, Germany}

\date{\today}

\begin{abstract}
The dynamics of an oscillated sessile drop of incompressible liquid with the focus on the contact line hysteresis is under theoretical consideration. The solid substrate is subject to transverse oscillations, which are assumed small amplitude and high frequency. The dynamic boundary condition that involves an ambiguous dependence of the contact angle on the contact line velocity is applied: the contact line starts to move only when the deviation of the contact angle exceeds a certain critical value. As a result, the stick-slip dynamics can be observed. The frequency response of surface oscillations on the substrate and at the pole of the drop are analyzed. It is shown that novel features such as the emergence of antiresonant frequency bands and nontrivial competition of different resonances are caused by contact line hysteresis.
\end{abstract}




\maketitle

\section{\label{sec:intro}Introduction}

Rapid development of microtechnologies over last decades has manifested great interest in theoretical aspects of contact line dynamics.
The ability to predict the motion of contact line and hence to control wetting processes becomes of paramount importance for applications.\cite{deGennes-85, Leger-Joanny-92, Rauscher-Dietrich-08} Despite noticeable progress in understanding the steady motion of the contact line, the unsteady motion remains significantly less explored and involves a number of important open questions. Of special interest is the role of contact angle hysteresis, which for unsteady motion may become a crucial feature in obtaining the proper picture of the contact line motion. In the present study, we address this issue in the context of oscillated sessile drop.
%

The dynamics of a drop on oscillated substrate has been considered for many years, see recent surveys in Refs.~\onlinecite{Lyubimov-Lyubimova-Shklyaev-06, Vukasinovic-Smith-Glezer-07}. Recent experimental studies have shown that the contact line hysteresis can lead to such nontrivial effects as stick-slip dynamics of the contact line\cite{Noblin-Buguin-Brochard-Wyart-04}, climbing motion over inclined substrate,\cite{Brunet-Eggers-Deegan-07} and motion over gradient\cite{Daniel-Chaudhury-02,Daniel-etal-04} and thermal gradient\cite{Mettu-Chaudhury-08} surfaces. These experimental observations raise a natural question about the role of the contact angle hysteresis, which is easy to pose but rather difficult to answer. In most theoretical studies, the contact angle hysteresis has been either completely neglected\cite{Lyubimov-Lyubimova-Shklyaev-06, Lyubimov-Lyubimova-Shklyaev-04, Alabuzhev-Lyubimov-07, Shklyaev-Straube-08} or treated in an oversimplified way, where the drop is similar to an oscillator with solid friction.\cite{Mettu-Chaudhury-08, Daniel-Chaudhury-deGennes-05, Buguin-Brochard-deGennes-06, Brochard-deGennes-07, Fleishman-Asscher-Urbakh-07} Although these solid-friction models reflect the qualitative picture of the stick-slip process, they do not provide satisfactory understanding of the phenomenon.

To obtain deeper insight into the physics of the stick-slip motion, a boundary condition suggested by L.~M.~Hocking\cite{Hocking-hyster-87} can be applied. This condition, which captures principal features of the contact line motion, involves an ambiguous dependence of the contact angle on the contact line velocity:
\begin{equation}
\frac {\partial {\zeta}}{\partial t} =
\left\{ \begin{array}{ll}
\Lambda(\chi-\chi_0), & \chi > \chi_0,  \\
0, & \left|\chi\right| \le \chi_0, \\
\Lambda(\chi+\chi_0), & \chi < -\chi_0.
\end{array}\right. \label{bc-hock-dim}
\end{equation}
\noindent Here, functions $\zeta$ and $\chi$ describe the deviations of the free surface and the contact angle from those in equilibrium, respectively, and $\chi_0$ is the critical value defining the contact angle hysteresis (see, e.g., Fig.~\ref{fig1}). The factor $\Lambda$, which has the dimension of velocity, characterizes interaction between the substrate and the liquid and is referred to as the wetting or the Hocking coefficient.

The particular case of $\chi_0= 0$, in which the contact line velocity $\partial\zeta/\partial t \propto \chi$, describes no contact angle hysteresis.\cite{Hocking-87} Different practically important situations can be addressed by changing $\Lambda$. In terms of the a corresponding dimensionless parameter, for instance, $\lambda$, as in relation \reff{dimless_param}, these situations range from the completely pinned contact line, $\lambda \to 0$ (the contact angle can change) to the opposite case of the fixed contact angle, $\lambda \to \infty$ (the contact line is freely moving).

In the present study, we consider the dynamics of a hemispherical drop on a normally oscillated solid substrate. We apply condition \reff{bc-hock-dim} without compromise and focus on the role of the contact line hysteresis, which has been recently measured in a similar setup.\cite{Noblin-Buguin-Brochard-Wyart-04} Based on this approach, we are not only able to quantitatively describe the stick-slip process but particularly to reveal a new interesting feature in the contact angle evolution. This finding is very much reminiscent of the experimental observations, which cannot be explained in terms of the previously suggested theoretical models.
The paper is outlined as follows. We start with the problem statement in Sec.~\ref{sec:problem}. Section~\ref{sec:solution} provides the description of the method we use to treat the problem. The obtained results are discussed in Sec.~\ref{sec:results} and summarized in Sec.~\ref{sec:concl}. \\

\section{Problem statement}\label{sec:problem}

\begin{figure}
\includegraphics[width=0.3\textwidth]{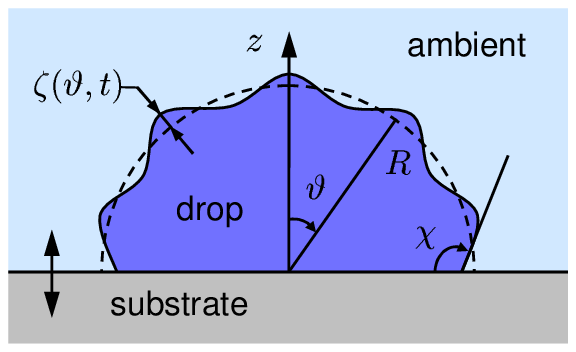}
\caption{Problem geometry. A hemispherical drop on the transversally oscillated substrate.} \label{fig1}
\end{figure}

Consider a sessile drop of incompressible liquid of density $\rho$ and kinematic viscosity $\nu$, see Fig.~\ref{fig1}. We are interested in the situation of a gaseous ambient, where its density is much smaller than that of the drop and therefore can be neglected. We assume that the solid substrate is subject to transverse oscillations with an amplitude $a$ and a frequency $\omega$. We admit that the drop is enough small so that its shape is hardly distorted by gravity. This assumption ensures that the equilibrium drop surface is hemispherical to very high accuracy, with radius $R$, and the equilibrium contact angle equals $\pi/2$.

The amplitude of external driving is considered small in the sense $\epsilon \equiv a/R \ll 1$ and the frequency of the substrate oscillations is high enough: $\omega R^2 \gg \nu $. At such frequencies, viscous boundary layers, which arise near the rigid plate and near the free surface, become very thin. In other words, the frequency restriction allows us to neglect viscous dissipation in the liquid, which ensures that the approximation of inviscid liquid is justified.\cite{Mei-Liu-73} On the other hand, the frequency $\omega$ is assumed comparable with the eigenfrequencies $\omega_n$ of shape oscillations for a spherical drop of radius $R$. For our consideration, only even eigenfrequencies are of interest, which spectrum is defined by the relation $\omega_n^2=2n(2n-1)(2n+2)\sigma/(\rho R^3)$, where $\sigma$ is the surface tension.

Because of symmetry, we use the spherical reference frame with the coordinates $r$, $\vartheta$, $\alpha$ and the origin at the center of drop and restrict our analysis by the axisymmetric problem. In the accepted approximations, the fluid motion is irrotational, which makes it convenient to introduce the velocity potential $\varphi$. As a result, the dynamics of the liquid is described by the Bernoulli equation and the incompressibility condition. Let $r=R+\zeta(\vartheta,t)$ be the instantaneous locus of the distorted free surface, see Fig.~\ref{fig1}. To make a comparison with the nonhysteresis study\cite{Lyubimov-Lyubimova-Shklyaev-06} simpler, we measure the time $t$, length $r$, velocity potential $\varphi$, the deviation of the pressure field $p$ from its equilibrium value and the surface deviation $\zeta$ in the scales of $\sqrt{ \rho R^3/ \sigma}$, $R$, $a \sqrt {\sigma/ \rho R}$, $a\sigma/ R^2$, and $a$, respectively.
As a result, the dimensionless boundary value problem is defined by (intermediate steps can be found in Ref.~\onlinecite{Lyubimov-Lyubimova-Shklyaev-06})
%
%
\begin{eqnarray}
&&p= - \frac{\partial \varphi}{\partial t} - \Omega^2 z \cos\Omega t, \quad \nabla^2 \varphi = 0, \label{gov-eqs}\\
&&\vartheta=\frac{\pi}{2}: \; \frac{\partial \varphi}{\partial \vartheta}=0,\label{bc-imperm}\\
&&r = 1: \ \frac{\partial \varphi}{\partial r} = \frac{\partial {\zeta}}{\partial t}, \ p + (\nabla^2_{\vartheta}+2)\zeta=0, \label{bc-free-surf}\\
&& r = 1, \; \vartheta=\frac{\pi}{2}: \;\frac {\partial {\zeta}}{\partial t} =-
\left\{ \begin{array}{ll}
\lambda(\gamma-\gamma_0), & \gamma > \gamma_0,  \\
0, & \left|\gamma\right| \le \gamma_0, \\
\lambda(\gamma+\gamma_0), & \gamma < -\gamma_0.
\end{array}\right. \label{bc-hock}
\end{eqnarray}
\noindent Here, the differential operator
\begin{equation}
\nonumber \nabla^2_{\vartheta}=\frac{1}{\sin\vartheta}
\frac{\partial}{\partial \vartheta} \left( \sin\vartheta
\frac{\partial\zeta}{\partial\vartheta}\right)
\end{equation}
\noindent and $\gamma=-(\partial\zeta/\partial\vartheta)|_{\vartheta=\pi/2}$ is the dimensionless deviation of the contact angle from its equilibrium value, which for the sake of brevity will be called the contact angle.

Boundary condition \reff{bc-imperm} ensures impermeability of the substrate for the liquid. The kinematic and the dynamics conditions at the free surface are presented by Eq.~\reff{bc-free-surf}. In contrast to the previous work,\cite{Lyubimov-Lyubimova-Shklyaev-06} we impose a more general Hocking condition at the line of contact of the three phases, as given by Eq.~\reff{bc-hock}. Thus, boundary value problem \reff{gov-eqs}-\reff{bc-hock} is similar to the previous study, except for this hysteretic condition.

The problem involves three dimensionless parameters
\begin{equation}
\label{dimless_param}\Omega^2=\frac{\rho \omega^2 R^3}{\sigma}, \quad \lambda=\chi\sqrt\frac{\rho R}{\sigma}, \quad \gamma_0=\frac{\chi_0 R}{a},
\end{equation}
which have the meaning of the squared external frequency rescaled with respect to the eigenfrequencies $\omega_n$, the wetting (or Hocking) parameter, and the critical value of contact angle, respectively.

As the frequencies $\omega$ and $\omega_n$ have been assumed comparable, the parameter $\Omega$ is finite. A similar argument refers to the parameter $\gamma_0$. We focus on the case of well polished substrate, which implies that the threshold value $\chi_0$ is small. Being the ratio of two small parameters, $\chi_0$ and $\epsilon$, the parameter $\gamma_0$ is treated as finite. This observation indicates that the contact-line hysteresis is expected to be non-negligible even for small-amplitude oscillations.

We emphasize that condition \reff{bc-hock} admits a number of particular cases. In the limiting case of perfectly polished surface, $\gamma_0 \to 0$, the boundary condition \reff{bc-hock} is reduced to its simplified modification.\cite{Lyubimov-Lyubimova-Shklyaev-06} Particularly, the contact angle remains fixed as $\lambda\to \infty$, whereas the opposite case, $\lambda \to 0$, describes the contact line pinned. We note that as in the latter case, the same dynamics refers to the limit of large values of $\gamma_0$.

Finally, we stress the nontriviality of our consideration. Although the amplitude of oscillations is considered small, $\epsilon \ll 1$, which allows us to linearize the governing equations and simplify the boundary conditions, the overall problem is {\em nonlinear}. The nonlinearity is ensured by posing the Hocking condition \reff{bc-hock} and comes into play through the parameter $\gamma_0 \sim a^{-1}$, therefore the solution is eventually amplitude dependent.

\section{\label{sec:solution}Method of solution}

To treat the formulated problem, we note that the velocity potential satisfies Laplace's equation \reff{gov-eqs} and therefore can be presented as a series in the Legendre polynomials. In view of impermeability condition \reff{bc-imperm}, only even terms are nonvaninshing in this expansion. We retain the terms regular at the origin and present the solutions for the velocity potential and the consistent solutions for the surface deviation and the pressure in the form
\begin{eqnarray}
\label{ansatz-zeta}\zeta&=&\sum_{n=1}^\infty C_{n}(t) P_{2n}(\theta),\\
\label{ansatz-varphi}\varphi&=&\varphi_0(t)+\sum_{n=1}^\infty \frac {\dot C_{n}(t)P_{2n}(\theta)r^{2n}}{2n},\\
\label{ansatz-p} p&=&p_0(t) - \sum_{n=1}^\infty \frac{\ddot C_{n}(t)P_{2n}\left( \theta \right)r^{2n}}{2n}- \Omega^2 z \cos\Omega t,
\end{eqnarray}
where $\theta\equiv\cos\vartheta$. Note that the summation for $\zeta$ starts from $n=1$. The term with $n=0$ is set to zero to ensure that the drop volume is conserved. The zeroth harmonics $\varphi_0(t)$ and $p_0(t)=-(\partial\varphi_0/\partial t)$ are nonvanishing spatially independent functions, which describe spatially uniform pulsations of the velocity potential and the pressure, respectively. The function $p_0$ is determined by the requirement of conservation of the drop volume, the term $\varphi_0$ is unimportant for the further analysis.

To distinguish between the time intervals of the contact line in motion from those in standstill, we next stick to the following notation.  Let $t=t_0$ be the time moment when the contact line stops to move and $t=t_1$ be the switching time when it proceeds with the motion again. We next obtain the solutions describing the dynamics at these time intervals separately and then show how to match the solutions.

\subsection{\label{subsec:pinned} Pinned contact line}

We now consider the time intervals characterized by small values of the contact angle, $|\gamma| < \gamma_0$. As it follows from the Hocking condition \reff{bc-hock}, the contact line remains fixed during this phase of evolution ($\lambda=0$) and we can use the solution obtained before.\cite{Lyubimov-Lyubimova-Shklyaev-06} The only point we have to care about is that in our situation the contact line is fixed not necessarily at $r=1$ but at a slightly different position characterized by $\zeta=\zeta_f \ne 0$, which is easy to take into account. As a result, the solution for $\zeta$ is presented as a superposition of the eigenmodes $\zeta_{0}^{(m)}$, a particular solution caused by the external force, $\zeta_p$, and the term $\propto \zeta_f$, which corrects the contact line position, respectively:
\begin{eqnarray}
\label{eq:zeta_t0}\nonumber \zeta(\theta,t)&=&\sum_{m=1}^\infty D_{m}\,\zeta_0^{(m)}(\theta)e^{i\omega_m(t-t_0)}+\zeta_p(\theta,t)\\
&&+ \zeta_f(1-2\theta).
\end{eqnarray}
\noindent The form of the last term is chosen such that the overall expression for $\zeta$ ensures the constant volume of the drop.
A similar ansatz for the velocity potential reads
\begin{eqnarray}
\label{eq:phi_t0} \varphi(r,\theta,t)&=&\varphi_0(t)+\sum_{m=1}^\infty D_{m}\,\varphi_0^{(m)}(r,\theta)e^{i\omega_m(t-t_0)} \nonumber\\
&&+ \varphi_p(r,\theta,t).
\end{eqnarray}
\noindent Here, $D_m$ are the complex amplitudes of the eigenoscillations to be determined as described in~\ref{subsec:matching} and $\omega_m$ are the eigenfrequencies for the drop with the pinned contact line. These eigenfrequencies are
defined as the roots of the transcendental equation
\begin{equation}
f(\omega,0)=0 \label{eq-omegam}
\end{equation}
\noindent with the function
\begin{equation}
f(x,\theta)=\sum_{n=1}^{\infty}\frac{\alpha_n \Omega_n^2 P_{2n}(\theta)}{\Omega_n^2-x^2}, \label{fun-f}
\end{equation}
\noindent where
\begin{eqnarray}
\alpha_{n}&=&-\frac{(4n+1)P_{2n}(0)}{(2n-1)(2n+2)}, \\
\Omega_n^2&=&2n(2n-1)(2n+2). \label{Omega2n}
\end{eqnarray}
\noindent The values $\Omega_n$ have the meaning of the dimensionless eigenfrequencies of even eigenmodes for the spherical drop oscillations. \\

The eigenfunctions for the problem with the fixed contact line ($\lambda=0$) are known to be~\cite{Lyubimov-Lyubimova-Shklyaev-06}
\begin{subequations}\label{zeta0-phi0}
\begin{eqnarray}
\label{zeta0m-fix} \zeta_0^{(m)}(\theta) = 2\sum_{n=1}^\infty n A_{mn}P_{2n}(\theta)=B_m f(\omega_m,\theta), \\
\label{eq:pot0_fix2} \varphi_0^{(m)}(r,\theta) = i \omega_m \sum_{n=0}^\infty A_{mn}P_{2n}(\theta)r^{2n}
\end{eqnarray}
\end{subequations}
\noindent with the coefficients
\begin{eqnarray}
\label{cff-Amn} &&A_{mn}=\frac{\alpha_{n}(2n-1)(2n+2)}{\Omega^2_{n}-\omega^2_{m}}B_m, \quad n \ge 0\\
\label{cff-Bm}&&B_m^{-2}=-\sum_{n=1}^\infty \frac{\alpha_{n}\Omega^2_{n}P_{2n}(0)}{\left(\Omega^2_{n}-\omega^2_{m}\right)^2}.
\end{eqnarray}
\noindent Here, we introduce the normalization condition and point out the orthogonality of the eigenfunctions
\begin{equation}
\label{eq:normBm} \int_0^1 \varphi_0^{(m)}(1,\theta)\zeta_0^{(k)}(\theta)d\theta=i\omega_m \delta_{mk}\\
\end{equation}
\noindent with $\delta=1$ for $m=k$ and $\delta=0$ otherwise.

For the problem of forced oscillations the solutions can be expressed as
\begin{subequations}\label{zetap-phip}
\begin{eqnarray}
\label{eq:zetap_fixc} \zeta_p(\theta,t)&=& {\rm Re}\left[\hat \zeta_p(\theta)  e^{i\Omega t}\right], \\
\label{eq:phip_fixc} \varphi_p(r,\theta,t)&=& {\rm Re}\left[\hat \varphi_p(r,\theta)  e^{i\Omega t}\right],
\end{eqnarray}
\end{subequations}
\noindent with the complex amplitudes
\begin{subequations}\label{zetap-phip-amp}
\begin{eqnarray}
\label{eq:zetap_fix} \hat \zeta_p&=&\Omega^2\left[\sum_{n=0}^\infty \frac { E_{n} P_{2n}(\theta)}{(2n-1)(2n+2)}+ g(\theta)\right],\\
\label{eq:phip_fix} \hat \varphi_p&=&i\Omega \sum_{n=0}^\infty E_{n}P_{2n}(\theta)r^{2n},
\end{eqnarray}
\end{subequations}
\noindent where
%
\begin{eqnarray}
g(\theta)&=&F \theta-\frac{1}{3}\left[1-\theta \ln \left(1+\theta \right)\right],\\
\label{cff-En} E_{n}&=&\Omega^2\alpha_{n}\left[\frac{1+(2n-1)(2n+2)F}{\Omega^2_{n}-\Omega^2}\right], \ n \ge 0, \\
\label{eq:F} F&=&-\frac{1} {f(\Omega, 0)}\sum_{n=1}^\infty  \frac {2n\alpha_{n}P_{2n}(0)}{\Omega^2_{n}-\Omega^2}.
\end{eqnarray}

We indicate that sum \reff{fun-f} included in relations \reff{eq-omegam}, \reff{zeta0m-fix} and \reff{eq:F} converges very slowly. From the computational point of view, its evaluation can be significantly improved if a more suitable form is used. By taking into account the expansion
\begin{equation}
\theta=\sum_{n=0}^{\infty}\alpha_{n}P_{2n}(\theta), \label{theta-expansion}
\end{equation}
sum \reff{fun-f} is presented in an alternative way
\begin{equation}
f(x,\theta)=\theta-\frac{1}{2}+ x^2 \sum_{n=1}^{\infty}\frac{\alpha_n P_{2n}(\theta)}{\Omega_n^2-x^2}, \label{fun-f-new}
\end{equation}
\noindent which compared with the original representation \reff{fun-f} provides much faster convergence. Note that a similar procedure can be applied to the sum in relation \reff{cff-Bm}.

\subsection{\label{subsec:mobile} Moving contact line}

We next deal with the time intervals of supercritical values of the contact angle, $|\gamma| > \gamma_0$, when the contact line is no longer fixed. We might build the solution in the form of series as we did in Sec.~\ref{subsec:pinned}. This way is however not worth implementing because the corresponding eigenvalue problem is not hermitian and hence no orthogonality condition as in \reff{eq:normBm} exists. As a result, this approach becomes computationally inefficient and provides no advantages any more. What we do instead, is addressing Eq.~\reff{gov-eqs} for the pressure, taken at $r=1$, which allows us to figure out how the contact line evolves. By using the dynamic boundary condition, Eq.~\reff{bc-free-surf}, we exclude the pressure and apply ansatz \reff{ansatz-zeta}. As a result, we obtain the inhomogeneous Legendre equation
\begin{eqnarray}
\label{eq:eqzeta}\nonumber \frac{\partial }{\partial \theta}\left[ \left( 1-\theta^2 \right) \frac{\partial \zeta}{\partial \theta}\right]&+&2\zeta = -p_0(t)+\Omega^2  \theta  \cos\Omega t\\
&&+ \sum_{n=1}^\infty \frac{\ddot C_{n}(t)P_{2n}(\theta)}{2n}.
\end{eqnarray}
%

The solution of this equation is given by
\begin{eqnarray}
\label{eq:zeta}\nonumber \zeta(\theta,t) &=& - \Omega^2 \cos\Omega t \sum_{n=1}^\infty \frac{\alpha_{n} P_{2n}(\theta)}{(2n-1)(2n+2)} \\
&-& \sum_{n=1}^\infty \frac{\ddot C_{n}(t) P_{2n}(\theta)}{2n(2n-1)(2n+2)} + \gamma\left(\theta-\frac{1}{2}\right),
\end{eqnarray}
\noindent where the term $\propto \gamma$ is the general solution of the homogeneous equation and the first two terms present a partial solution of the inhomogeneous equation. We note that the integration ``constant,'' $\gamma=\gamma(t)$, satisfies the definition of the contact angle
\begin{equation}
\label{def-gamma} \left.\frac{\partial \zeta}{\partial \theta}\right|_{\theta=0}\equiv \gamma.
\end{equation}
\noindent For this reason, the time-dependent function $\gamma(t)$ is referred to as the contact angle.

By making comparison of expressions \reff{ansatz-zeta} and \reff{eq:zeta} and using relations \reff{Omega2n} and \reff{theta-expansion}, we derive a set of ordinary differential equations for the expansion coefficients $C_{n}$
\begin{equation}
\label{eq:C2n_eq} \ddot C_{n} + \Omega_{n}^2C_{n} =
\Omega_{n}^2 \alpha_{n} \gamma - 2n \Omega^2 \alpha_{n} \cos \Omega t,
\end{equation}
\noindent which are coupled to each other through $\gamma(t)$. As the function $\gamma(t)$ is unknown, an additional relation is required to close the system. The formulation of the problem is completed by rewriting the Hocking condition \reff{bc-hock}, which yields
\begin{equation}
\label{eq-gamma(t)} \gamma(t) =
\left\{ \begin{array}{ll}
S(t)+\gamma_0, & \gamma > \gamma_0,  \\
S(t)-\gamma_0, & \gamma < -\gamma_0,
\end{array}\right.
\end{equation}
\noindent with the auxiliary function
\begin{equation}
S(t)=\frac{1}{\lambda} \sum_{n=1}^\infty \dot C_{n}(t) P_{2n}(0). \label{fun-S}
\end{equation}

Thus, at the stage of evolution with supercritical contact angles, $|\gamma|>\gamma_0$, we numerically solve the system of ordinary inhomogeneous differential equations (\ref{eq:C2n_eq}) together with algebraic coupling relation \reff{eq-gamma(t)}.

\subsection{\label{subsec:matching} Matching the solutions}

To obtain the solution of the full problem, we have to match the solutions obtained in Secs.~\ref{subsec:pinned} and \ref{subsec:mobile}. The regime with the motionless contact line, which is characterized by subcritical values of the contact angle, $|\gamma|<\gamma_0$, is described by expressions \reff{eq:zeta_t0} and \reff{eq:phi_t0} with the unknown complex-valued coefficients $D_m$. At supercritical values of the contact angle, $|\gamma|>\gamma_0$, the contact line keeps moving. At this time interval, we treat Eqs.~\reff{eq:C2n_eq}-\reff{fun-S} numerically, in terms of functions $C_n(t)$ and $\dot C_n(t)$. Next we provide relations between the coefficients $D_m$ and $C_n$, $\dot C_n$ valid at $t=t_0$ and $t=t_1$, at which the regimes are switched. At these moments, the contact angle reaches its critical value, $\gamma=\gamma_0$, and the two solutions coincide.

Consider first the moment $t=t_0$, when the switchover from the regime with the moving contact line to the one with the pinned contact line occurs. Given the values $C_n(t_0)$ and $\dot C_n(t_0)$ obtained at the previous phase of motion, we have to determine coefficients $D_m$. We multiply Eqs.~(\ref{eq:zeta_t0}) and (\ref{eq:phi_t0}) by $\varphi_0^{(m)}(1,\theta)$ and $\zeta_0^{(m)}(\theta)$, respectively, take into consideration integral condition (\ref{eq:normBm}), and obtain the real, $D_m^{(r)}$, and imaginary, $D_m^{(i)}$, parts of $D_m$
\begin{eqnarray}
\label{eq:rD_m} D_m^{(r)} &=& \sum_{n=1}^\infty \frac{A_{mn}}{4n+1}\left( C_{n}- 2nE_{n} \cos \Omega t_0 \right) - \frac{\zeta_f B_m}{\omega^2_{m}},\nonumber\\
\label{eq:iD_m} D_m^{(i)} &=& -\frac{1}{\omega_m}\sum_{n=1}^\infty \frac{A_{mn}}{4n+1}\left(\dot C_{n}+2n\Omega E_{n} \sin \Omega t_0 \right). \nonumber
\end{eqnarray}
\noindent Here, the values $\omega_m$ are determined as the roots of Eq.~\reff{eq-omegam} and the coefficients $A_{mn}$, $B_m$, and $E_n$ are given by relations \reff{cff-Amn}, \reff{cff-Bm}, and \reff{cff-En}, respectively.

This transformation identifies solutions \reff{eq:zeta_t0} and \reff{eq:phi_t0} uniquely and allows us to determine the moment of the backward switchover, $t=t_1$, when the contact line starts to move again. To find out this moment, we numerically solve the algebraic equation with respect to $t_1$
\begin{equation}
\gamma(t_1)-\gamma_0=0.
\end{equation}
\noindent To evaluate $\gamma$ we use Eq.~\reff{def-gamma} with the solution for $\zeta$, Eq.~\reff{eq:zeta_t0}, where the sum in expression \reff{zeta0m-fix} should be taken in the form \reff{fun-f-new}. As a result, we obtain
\begin{equation}
\gamma(t)=\Omega^2 F \cos \Omega t-2\zeta_f +\sum_{m=1}^\infty B_m d_m^{(r)}(t).
\end{equation}
\noindent Here, we have introduced a complex valued function of time $d_m(t) \equiv D_m \exp[i\omega_m(t-t_0)]$ with the real and imaginary parts denoted as $d_m^{(r)}$ and $d_m^{(i)}$, respectively. Thus, having obtained the value $t_1$, we are ready to proceed to the next situation. \\

We now turn to the consideration of the moment $t=t_1$, when the motionless contact line starts to move. Before we treat Eqs.~\reff{eq:C2n_eq}-\reff{fun-S}, we need to evaluate initial values $C_n(t_1)$ and $\dot C_n(t_1)$.
We equate solutions (\ref{eq:zeta_t0}) and (\ref{eq:phi_t0}) to those in relations \reff{ansatz-zeta} and \reff{ansatz-varphi}, which are taken at $r=1$ and $t=t_1$. All the terms are presented as series in the Legendre polynomials. This can be done with the aid of expressions \reff{zeta0-phi0}, \reff{zetap-phip}, and \reff{zetap-phip-amp}. The coefficients on the left and right hand sides must be equal, which yields
\begin{eqnarray}
\label{eq:zeta_t1C_start} \nonumber C_{n} &=& 2n\left[ \sum_{m=1}^\infty d_m^{(r)}(t_1) A_{mn} + E_n\cos \Omega t_1 \right] - 2\zeta_f \alpha_{n},\\
\label{eq:dotC_start} \nonumber \dot C_{n} &=&-2n\left[ \sum_{m=1}^\infty \omega_m d_m^{(i)}(t_1) A_{mn} +\Omega E_n \sin \Omega t_1 \right].
\end{eqnarray}

Thus, we start from these initial values and solve Eqs.~\reff{eq:C2n_eq}-\reff{fun-S} until the condition $|\gamma|<\gamma_0$ is fulfilled. Suppose that $\gamma$ crosses the critical value $\gamma_0$ between the time steps $k$ and $(k+1)$. We need to find the moment $t=t_0$ with the accuracy consistent with the integration scheme and evaluate the values $C_n(t_0)$ and $\dot C_n(t_0)$. To estimate $t_0$ we implement an elegant method suggested by H\'enon.\cite{Henon-82} The idea is to introduce a characteristics that changes its sign while $\gamma$ is crossing the value $\gamma_0$. A suitable quantity satisfying this requirement can be the function $S(t)$, see relation \reff{fun-S}, which turns to zero at $\gamma(t_0)=\gamma_0$. Thus, $t_0$ as well as the values $C_n(t_0)$ and $\dot C_n(t_0)$ are obtained by making one additional corrective integration step $S(t_{k})$ from the values $C_{n}(t_{k})$ and $\dot C_n(t_{k})$, which are available at time step $k$.

To have this idea implemented, we proceed from Eqs.~\reff{eq:C2n_eq} to the system of differential equations with the new independent variable $S$ and the time $t=t(S)$ as an additional variable
\begin{subequations}\label{henon-ode}
\begin{eqnarray}
\label{eq:our_eno_sys} \frac{dC_{n}}{dS}&=&\frac{\dot C_{n}}{H},\\
\frac{d\dot C_{n}}{dS}&=&\frac{ \Omega^2_{n}\left(\alpha_{n}\gamma - C_{n} \right) -2n\Omega^2\alpha_{n}\cos \Omega t}{H},\\
\frac{dt}{dS}&=&\frac{1}{H}.
\end{eqnarray}
\end{subequations}
\noindent Here, we have introduced the H\'enon function
\begin{equation}
\label{fun-H} H=\frac{dS}{dt}=\frac{1}{\lambda} \sum_{n=1}^{\infty} \ddot C_{n} P_{2n}(0),
\end{equation}
\noindent where $\ddot C_n$ can be expressed from Eqs.~(\ref{eq:C2n_eq}).

Finally, we integrate Eqs.~\reff{henon-ode} along with relation \reff{eq-gamma(t)} until $S$ changes its sign. We note that while making the regular integration steps one sets $H=1$. The corrective integration with the step $S(t_{k})$ is made with $H$ in the form \reff{fun-H}, which after the correction corresponds to $S=0$ or equivalently to $t=t_1$ and hence provides the required $C_n(t_1)$ and $\dot C_n(t_1)$.

Thus, we started with the consideration of the moment $t=t_0$, provided the way of proceeding to the moment $t=t_1$ and then to the next moment $t=t_0$. To this end, we have obtained the solution over half of the period and the described matching procedure can be successively repeated to obtain the solution at longer times.

In numerical calculations, infinite number of eigenmodes in Eqs.~\reff{eq:zeta_t0} and \reff{eq:phi_t0} was
truncated to retain $M$ terms. The presented results were calculated for $M=10$. The control computations with the number
of the eigenmodes with up to $M=20$ have indicated no change in the results. However, a further increase of $M$ leads to the emergence of unphysical oscillations. The number $N$ of Legendre harmonics retained in Eqs.~\reff{ansatz-zeta}-\reff{ansatz-p} and Eqs.~\reff{zeta0-phi0} and \reff{zetap-phip-amp} was chosen to be $100$ in most of calculations. In order to check the accuracy of calculations we performed a number of tests with $N=150$ and $N=200$, which gave very close results.

\section{\label{sec:results}Results and discussion}

We start our discussion by recalling the fact\cite{Hocking-87} that despite the neglected mechanism of viscous dissipation, the Hocking condition itself is dissipative. Exceptions are the particular case of the pinned contact line ($\lambda \to 0$) and the contact line freely moving ($\lambda \to \infty$) over the perfectly polished ($\gamma_0 \to 0$) substrate. Because the system under consideration is generally dissipative, any initial state approaches the terminal oscillatory state after a certain transient. In other words, any phase trajectory is landing at a limit cycle. Because in the case $\gamma_0 \to 0$ and $\lambda=O(1)$ the decay rate is comparable with the frequency of oscillation,\cite{Lyubimov-Lyubimova-Shklyaev-06} the transient time is estimated to be a few periods of oscillations. For these reasons, we are mostly interested in the properties of the steady-state oscillations.

\begin{figure}
\includegraphics[width=0.48\textwidth]{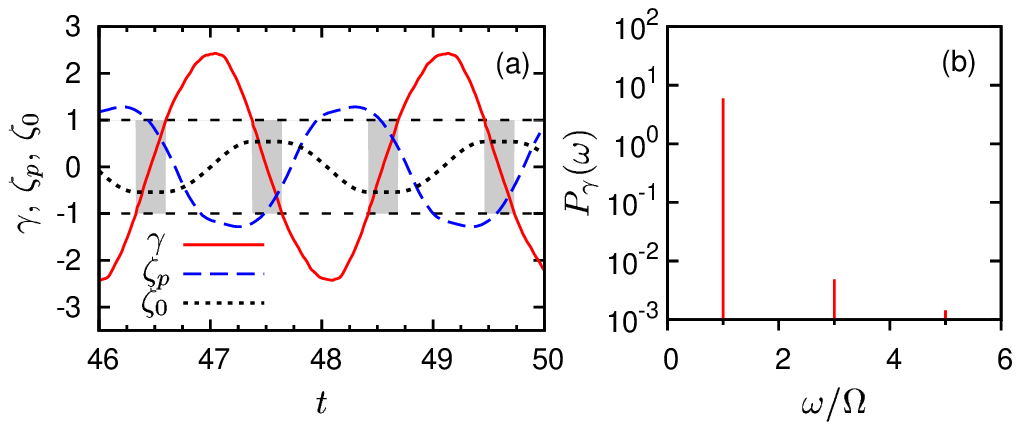}
\caption{\label{fig2}Characteristics of the steady-state oscillations at $\lambda=1$, $\gamma_0=1$, and $\Omega=3$. Panel (a): Evolutions of the contact angle, $\gamma(t)$, deviations of the free surface at the pole, $\zeta_p(t)$, and at the contact line, $\zeta_0(t)$. The horizontal dashed lines are the lines $\gamma=\pm\gamma_0$. The filled areas display the time intervals of the subcritical contact angles, $|\gamma|<\gamma_0$. Panel (b): The Fourier power spectrum evaluated for $\gamma(t)$.}
\end{figure}

To get an impression about the dynamics, we look at the steady-state oscillations of the contact angle, see Fig.~\ref{fig2}(a). When one sees the evolution of $\gamma$, it might be thought of simple linear oscillations. However, despite a relatively simple form of the observed signal, the oscillations are nonlinear, which is well seen from the Fourier spectrum of the signal, Fig.~\ref{fig2}(b). We note that although the driving frequency dominates, a few higher harmonics are nonvanishing. Another feature one can readily notice is the absence of the even harmonics, which reflects the fact that the response is presented by an antisymmetric function. This antisymmetry of the steady-state oscillations can be seen directly from our mathematical model. Indeed, in the terminal state the system is invariant with respect to the transformation
\begin{equation}\label{invar}
t \to t+\frac{\pi}{\Omega}, \ \zeta \to -\zeta, \ \varphi \to -\varphi,
\end{equation}
\noindent which is easily understood if one takes into account two circumstances. First, the governing equations and boundary conditions, Eqs.~\reff{gov-eqs}-\reff{bc-hock}, are linear, if considered at intervals with the fixed and moving contact line separately. Only the periodic switching between these regimes makes the problem nonlinear. We also note that the problem, Eqs.~\reff{gov-eqs}-\reff{bc-hock}, is inhomogeneous. However, the inhomogeneity $\propto \cos\Omega t$ changes its sign under time transformation \reff{invar}, as required. Second, we take the same threshold value $\gamma_0$ used for the advancing and receding motion of the contact line. Our numerical tests with $\gamma_0$ and $\gamma_1\ne \gamma_0$ for the thresholds of the advancing and receding motion, respectively, have confirmed this statement. For the distinct threshold values, we detect nonvanishing even harmonics, which contributions to the power spectrum become more pronounced as $\gamma_0$ and $\gamma_1$ become more distinct. In the opposite case of $\gamma_1 \to \gamma_0$, the even harmonics die out and we come back to the perfect antisymmetry, as in relation \reff{invar}.

Along with the contact angle, we measure the deviations of the free surface from its equilibrium position at the pole of the hemisphere, $\zeta_p(t)\equiv \zeta(\theta=1,t)$ and on the substrate $\zeta_0(t)\equiv \zeta(\theta=0,t)$, Fig.~\ref{fig2}(a). The latter characteristics shows the dynamics of the contact line. We see that the evolution of the system consists of two interchanging regimes. During the time intervals characterized by supercritical values of the contact angle, $|\gamma(t)|>\gamma_0$, the contact line keeps sliding over the substrate. This motion takes place until $\gamma$ enters the subcritical domain, $-\gamma_0<\gamma(t)<\gamma_0$, where the contact line becomes ``frozen.'' In Fig.~\ref{fig2}(a), the time intervals of the contact line being fixed are presented as the gray-filled areas. As we clearly see, $\zeta_0(t)={\rm const}$ here, whereas other characteristics are changing. The contact line remains fixed until the contact angle is outside the subcritical domain. After that, the contact line proceeds to move again, etc. Thus, the contact line dynamics corresponds to the periodic sliding interrupted by the intervals of being completely frozen, or, in other words, to stick-slip motion.

\begin{figure}
\includegraphics[width=0.48\textwidth]{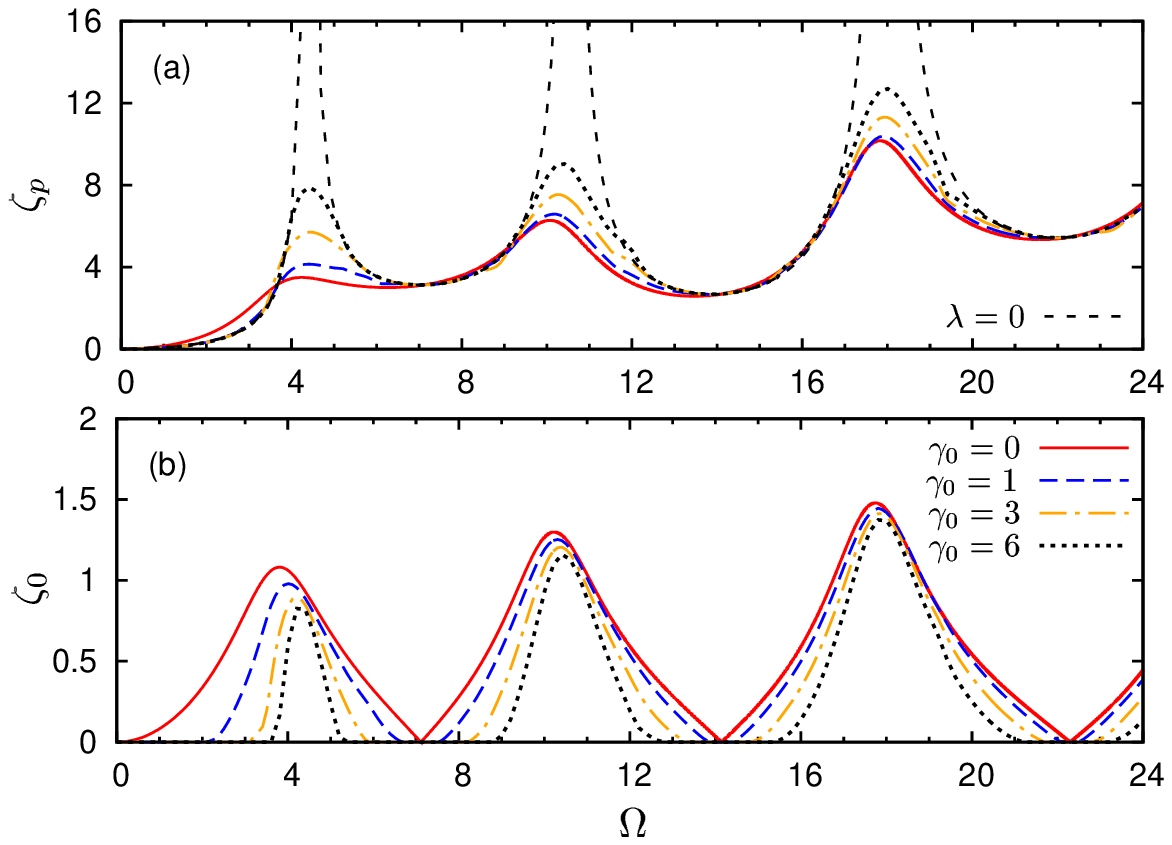}
\caption{\label{fig3} Amplitude-frequency response at $\lambda=1$ and different values of $\gamma_0$. Panel (a): Surface deviation at the pole, $\zeta_p$. Panel (b): Surface deviation on the substrate, $\zeta_0$. }
\end{figure}

We now keep the value of the wetting parameter fixed, $\lambda=1$, and analyze how the amplitudes of $\zeta_p$ and $\zeta_0$ change while varying the external frequency $\Omega$ and the critical contact angle, $\gamma_0$. The corresponding dependencies are depicted in Fig.~\ref{fig3}. As we see from the form of these response characteristics, the system demonstrates well pronounced resonances. Because the contact line motion with a finite value of the wetting parameter $\lambda$ is dissipative, the resonant amplitudes remain bounded. In Fig.~\ref{fig3}(a) we also present the nondissipative limiting case of $\lambda=0$.

We next pay attention to the dependence on $\gamma_0$. The partial case of $\gamma_0=0$ corresponds to no hysteresis
so that at $\lambda=1$ the contact line keeps moving all the time.
With the increase of $\gamma_0$, the part of period with the moving contact line becomes less and is gradually replaced with the regime with the fixed contact line. At large values of $\gamma_0$, the contact line remains fixed for the most part of period, which becomes equivalent to the case of small $\lambda$. In other words, the dynamics with the fixed contact line dominates.
At $\lambda = 0$, the contact line is pinned, $\zeta_0=0$, and the oscillations are no longer dissipative, which results in the divergence of resonant amplitudes of $\zeta_p$. We note that the case of $\lambda=1$ is characterized by the resonant frequencies close to $\omega_m$ for all $\gamma_0$, where $\omega_m$ are the eigenfrequencies at $\lambda=0$, see Eqs.~\reff{eq-omegam} and \reff{fun-f}. As a consequence, the amplitudes of oscillations at the pole, $\zeta_p$, are significantly higher than those on the contact line, $\zeta_0$.

Let us now point out another generic feature caused by the contact line hysteresis. We start with the limiting case of no hysteresis, $\gamma_0=0$. We recall that in this case,\cite{Lyubimov-Lyubimova-Shklyaev-06} the contact line remains fixed, $\zeta_0=0$, at certain values of the driving frequency, $\Omega=\Omega_{ar}$, and any value of the wetting parameter, $\lambda$. For this reason, the values $\Omega_{ar}$ are referred to as {\em antiresonant}. Such frequencies are well recognized in Fig.~\ref{fig3}(b).
%
%
As becomes clear from the figure, the contact line hysteresis, $\gamma_0 \ne 0$, transforms the discrete number of antiresonant points into antiresonant {\em bands} of finite width, Fig.~\ref{fig3}(b). With the growth of $\gamma_0$, the islands of stick-slip dynamics become narrower, whereas the regions of behavior with the completely fixed contact line widen.

\begin{figure}
\includegraphics[width=0.48\textwidth]{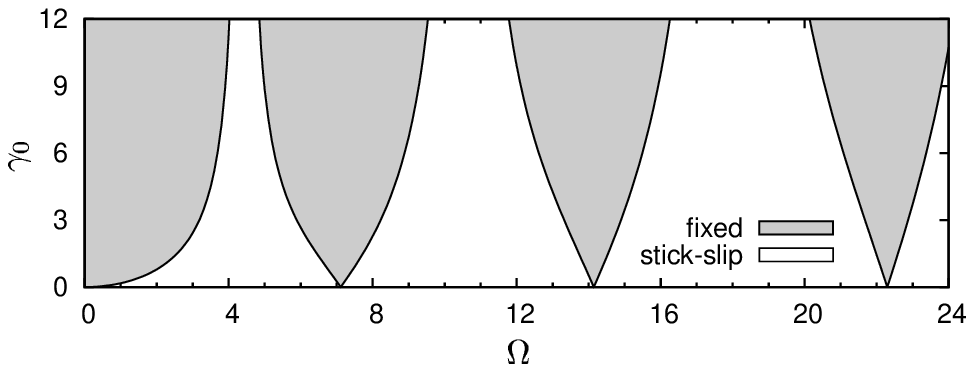}
\caption{\label{fig4} The diagram of contact line motion on the plane ($\Omega$, $\gamma_0$). The solid lines are defined by the condition $\Gamma(\Omega)=\gamma_0$ and separate the domains of oscillations with the fixed contact line ($\Gamma<\gamma_0$, in gray) and with the contact line moving in the stick-slip regime ($\Gamma>\gamma_0$). Courtesy of S.~Shklyaev.
%
}
\end{figure}

We emphasize that the width of the antiresonant bands is determined solely by the value of $\gamma_0$ and is independent of $\lambda$, which is explained as follows. The dynamics at frequencies within the antiresonant band corresponds to the oscillations with the fixed contact line, as if $\lambda=0$. It is clear from Eq.~\reff{bc-hock}, that here we have $\Gamma(\Omega)<\gamma_0$ with $\Gamma=\max \gamma(t)$, whereas outside the domain of antiresonant behavior the opposite equality holds, $\Gamma(\Omega)>\gamma_0$. Hence, the border between the domains of stick-slip dynamics and behavior with the fixed contact line is defined by the equality $\Gamma(\Omega)=\gamma_0$ or even much simpler: $\Gamma_0(\Omega)=\gamma_0$, where $\Gamma_0=\Gamma|_{\lambda=0}$. As we see, the question about the width of the antiresonant bands can be efficiently answered within the nonhysteretic model.\cite{Lyubimov-Lyubimova-Shklyaev-06} A corresponding diagram is shown in Fig.~\ref{fig4}.

The next question to answer is if one might expect any significant difference in the dynamics for other values of the wetting parameter.
The case of small values of $\lambda$ is out of interest because the dynamics becomes very similar to the case of the pinned contact line, $\lambda \to 0$. Our numerical tests show that this limit is practically reached at $\lambda=1/2$. As a result, the variation of $\gamma_0$ demonstrates almost no significant changes and hence the case of $\lambda <1$ brings basically nothing new.

Much more promising is the opposite situation, $\lambda \gg 1$. In the nonhysteretic ($\gamma_0 \ne 0$) case, the contact line is not fixed and the sliding motion is predominant. In contrast to the case of $\lambda \ll1$, interaction of the drop with the substrate is weakened. The eigenfrequencies become close to the eigenfrequencies of the even modes for a spherical drop, $\Omega_n$. In the case of hysteresis, $\gamma_0 \ne 0$, the system is switched between two weakly dissipative kinds of oscillations. As we have seen for the case of $\lambda=1$, the stage of evolution with the fixed contact line is characterized by the resonant frequencies $\omega_m$. For the stage of sliding contact line, the resonances are found at $\Omega_n<\omega_n$. Thus, a competition of the qualitatively different resonances is expected for the stick-slip motion.

\begin{figure}
\includegraphics[width=0.48\textwidth]{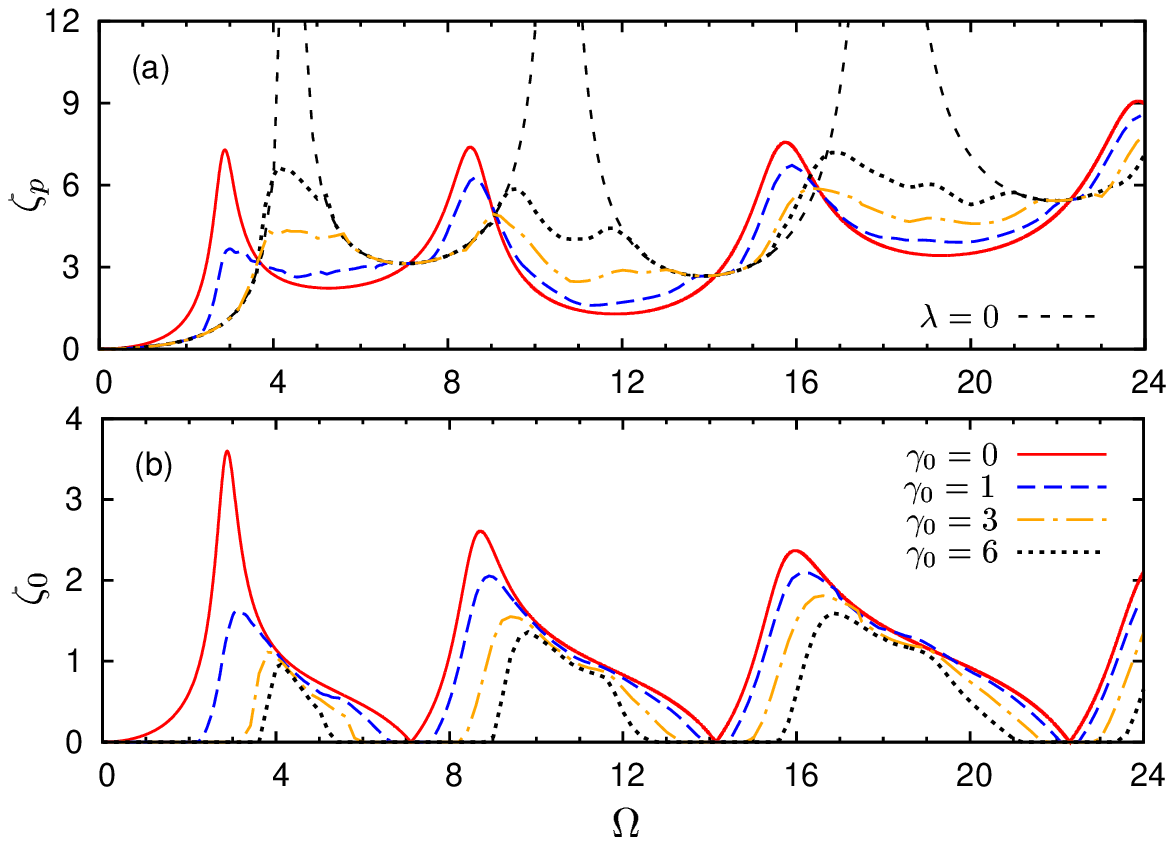}
\caption{\label{fig5} Amplitude-frequency response at $\lambda=5$ and different values of $\gamma_0$. Panel (a): Surface deviation at the pole, $\zeta_p$. Panel (b): Surface deviation on the substrate, $\zeta_0$.}
\end{figure}

Our computations indicate that the described scenario with $\lambda \gg 1$ can be observed already at $\lambda=5$. The corresponding response characteristics $\zeta_p$ and $\zeta_0$ are shown in Fig.~\ref{fig5}, where the competition of pairs of neighboring resonances for $\zeta_p$ is well seen. As in the case of $\lambda=1$, the growth of $\gamma_0$ demonstrates convergence to the resonant curve corresponding to the fixed contact line, $\lambda=0$. The transition is however nontrivial. In contrast to Fig.~\ref{fig3}(a), the curve associated with $\gamma_0=0$ has resonant peaks at $\Omega=\Omega_n$. As the parameter $\gamma_0$ is increased, the peaks do not simply shift from $\Omega=\Omega_n$ toward the values $\Omega=\omega_n$. This transition is accompanied by the emergence of intermediate local maxima. Another distinction is that the amplitudes of resonant peaks change now nonmonotonically with the increase of $\gamma_0$. For instance, consider the amplitude of $\zeta_p$ in a vicinity of the first resonance, $\Omega\in(2,6)$. At $\gamma_0=0$, we have the maximum value $\zeta_p\approx 7.30$. With the growth of $\gamma_0$, the resonant amplitude starts to decrease and approaches its minimal value $\zeta_p\approx 3.09$ at $\gamma_0=1.45$. The further increase of $\gamma_0$ leads to the growth and then divergence of the resonant amplitude, as in the case of $\lambda=0$.

The dependence of the amplitude $\zeta_0$ on $\gamma_0$ and $\Omega$ is qualitatively the same as described for $\lambda=1$ and is in agreement with the diagram of contact line motion, Fig.~\ref{fig4}. The amplitudes of $\zeta_0$ have, however, higher values because of weaker dissipation than those at $\lambda=1$. \\

\begin{figure}
\includegraphics[width=0.48\textwidth]{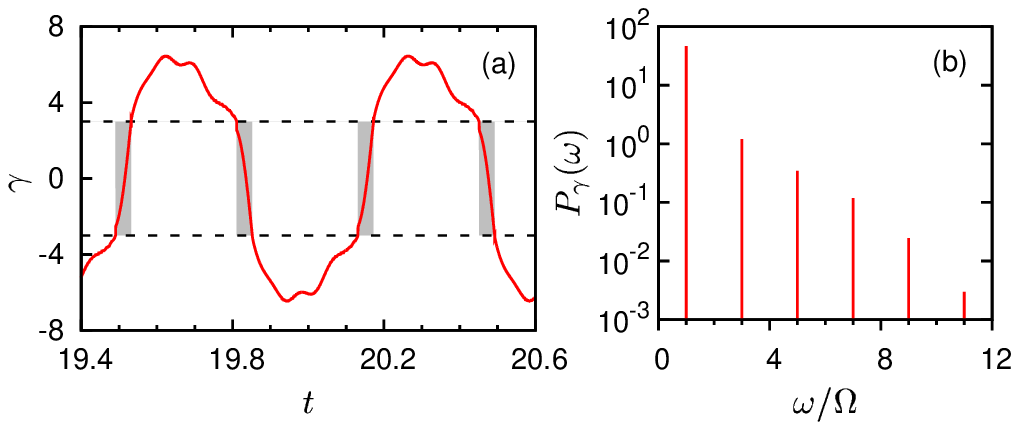}
\caption{\label{fig6} Characteristics of the steady-state oscillations at $\lambda=5$, $\gamma_0=3$, and $\Omega=9.8$. Panel (a): Evolution of the contact angle, $\gamma(t)$. The horizontal dashed lines are the lines $\gamma=\pm\gamma_0$. The filled areas display the time intervals of the subcritical contact angles, $|\gamma|<\gamma_0$. Panel (b): The Fourier power spectrum evaluated for $\gamma(t)$.}
\end{figure}

We next examine the evolution of $\gamma$ and its Fourier spectrum
evaluated at $\lambda=5$ and $\gamma_0=3$, see Fig.~\ref{fig6}. As we see, the dependence $\gamma(t)$ looks not only more complicated, but qualitatively different in comparison with that given in Fig.~\ref{fig2}. To avoid any confusion, we hereafter stick to the following convention. We focus on half the period of the signal $\gamma(t)$ such that $\gamma>0$. Note that in the case discussed in Fig.~\ref{fig2}(a) we see a single maximum. If we now go back to Fig.~\ref{fig6}(a) we detect the birth of the second {\em local} maximum, the origin of which is discussed in a few lines. As a result, the power spectrum becomes wider than in Fig.~\ref{fig2}(b) and the contribution of higher harmonics is stronger.
It is important to indicate that a very similar feature has been recently observed experimentally, see Fig.~11, Ref.~\onlinecite{Noblin-Buguin-Brochard-Wyart-04}. Along with the experimental study, the authors have suggested a simple theoretical model. Although their model is able to qualitatively explain the existence of the stick-slip motion, it fails to reproduce the two-maxima feature in the evolution of the contact angle, $\gamma(t)$. Although the considered problems are not exactly the same, the advantage of our approach comes into play. Our model allows us not only to describe the stick-slip motion itself but also to capture the subtle feature of non-single maximum in the evolution of contact angle.

\begin{figure}
\includegraphics[width=0.48\textwidth]{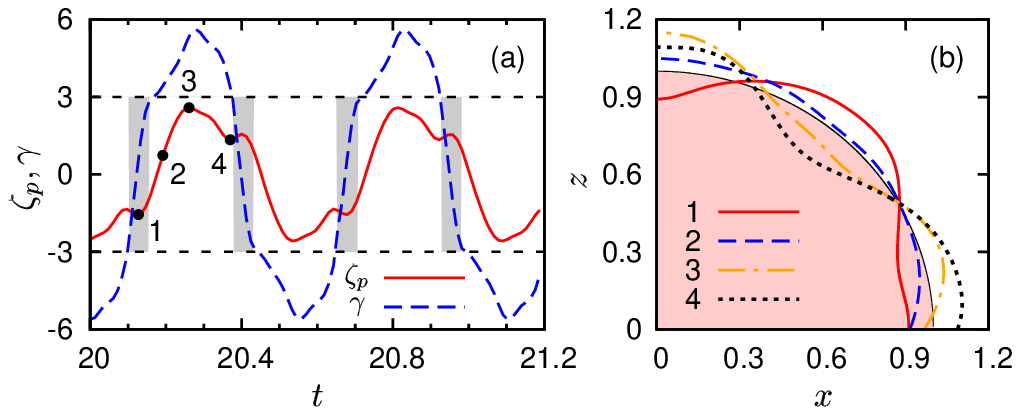}
\caption{\label{fig7} Characteristics of the steady-state oscillations at $\lambda=5$, $\gamma_0=3$, and $\Omega=11.4$. Panel (a): Evolutions of the free surface deviation at the pole, $\zeta_p(t)$ and the contact angle, $\gamma(t)$. The horizontal dashed lines are the lines $\gamma=\pm\gamma_0$. The filled areas display the time intervals of the subcritical contact angles, $|\gamma|<\gamma_0$. Panel (b): Profiles of the free surface shown at the consecutive moments of time as indicated by points $1$, $2$, $3$, and $4$ in panel (a).}
\end{figure}

To get a deeper insight into the two maxima phenomenon, we provide Fig.~\ref{fig7} evaluated at $\lambda=5$, $\gamma_0=3$, and $\Omega=11.4$. We now consider half the period of $\zeta_p(t)$ with $\zeta_p>0$. Figure~\ref{fig7}(b) additionally presents the profiles of the drop at different times as indicated with circles in Fig.~\ref{fig7}(a).

As we see in Fig.~\ref{fig7}(a), the dependence $\zeta_p(t)$ possesses two local maxima. One maximum, which is similarly present for $\zeta_p(t)$ in Fig.~\ref{fig2}, concerns the stage of oscillations with the fixed contact line. The second maximum is new, it relates to the stage of sliding contact line. Note that each stage of motion is characterized by its own resonance, which are competing as we discussed for the case of $\lambda=5$, Fig.~\ref{fig5}. At $\gamma_0=0$, the motion with the fixed contact line is characterized by the resonances at the frequencies $\omega_m$, whereas for the slip motion the resonances at $\Omega_n$ become important. At $\Omega=11.4$ and $\gamma_0=0$, the closest eigenfrequencies are $\Omega_2=8.49$ (fixed contact angle) and $\omega_2=10.6$ (fixed contact line). If we look how these resonant values change as $\gamma_0$ is increased, we find that at $\gamma_0=3$ the resonances take place at $\tilde \Omega_2\approx 9.10$ and $\tilde \omega_2 \approx 12.0$ and the value $\Omega=11.4$ is well in between and close to both of them. This explanation may also reveal the reason behind the two maxima as in Fig.~\ref{fig6}. Despite both those maxima are found during the slip motion, they are caused by the competing resonances of different nature, as we described.

\section{\label{sec:concl}Conclusions}

We have considered the dynamics of an oscillated sessile drop of incompressible liquid and focused on the contact line hysteresis. The solid substrate is subject to transverse oscillations, which are assumed small amplitude and high frequency. We admit that the drop is so small that its shape is not distorted by gravity and hence the equilibrium drop surface is hemispherical and the equilibrium contact angle equals $\pi/2$. To take into consideration the contact line hysteresis, the boundary condition suggested by L.~M.~Hocking is applied, see Eq.~\reff{bc-hock-dim}. This boundary condition involves an ambiguous dependence of the contact angle on the contact line velocity. More precisely, the contact line starts to move only when the deviation of the contact angle exceeds a certain critical value. As a result, the stick-slip dynamics can be observed: the system is periodically switched between the states with the sliding and the completely fixed contact line.

The solution of the boundary value problem is presented as series in the Legendre polynomials. Technically, the problem is treated by building two separate solutions valid at subcritical and supercritical values of the contact angle. These solutions are different and correspond to oscillations with the completely fixed and the moving contact line, respectively. For the fixed contact line, the problem admits an analytical solution obtained earlier.\cite{Lyubimov-Lyubimova-Shklyaev-06} In the situation with the moving contact line, a set of ordinary differential equations is obtained for expansion coefficients, which are integrated numerically. At the critical values of the contact angle, the matching of the different solutions is performed. This procedure allows one to obtain the solution of the formulated problem at any moment of time.

Because of dissipative nature of the Hocking condition, the regime with steady nonlinear oscillations is reached. We have measured the deviations of the free surface on the substrate and analyzed the frequency response at different values of the wetting parameter, $\lambda$, and the critical contact angle, $\gamma_0$. It is known that in the nonhysteretic limit, $\gamma_0=0$, no contact line motion exists at certain frequencies $\Omega=\Omega_{ar}$, which are independent of the wetting parameter. For this reason, the values $\Omega_{ar}$ are referred to as antiresonant frequencies. We have shown that the contact line hysteresis, when $\gamma_0\ne0$, transforms this discrete number of $\Omega_{ar}$ into antiresonant frequency bands of finite width. With the growth of $\gamma_0$, the parameter domains of the stick-slip dynamics become narrower, whereas the one with the completely fixed contact line grows.

We have analyzed similar frequency response for the deviation of the free surface at the pole of the drop. Here, at relatively small values of the wetting parameter, $\lambda$, resonant amplification of oscillations is found at frequencies $\Omega \simeq \omega_n$ for all $\gamma_0$, where $\omega_n$ are the eigenfrequencies of the problem with the pinned contact line, $\lambda=0$.

At higher values of $\lambda$, the interaction with the substrate is weakened. In the case of no hysteresis, $\gamma_0=0$, the eigenfrequencies are close to the eigenfrequencies of the even modes for a spherical drop, $\Omega_n$. We have demonstrated that the contact line hysteresis leads to a nontrivial shift of resonant frequencies from $\Omega\simeq\Omega_n$ to $\omega_n$ as $\gamma_0 \to \infty$. For moderate values $\gamma_0\simeq O(1)$, the switching between two weakly dissipative kinds of oscillations: with the sliding and the completely fixed contact line. These stages of stick-slip motion are characterized by the resonant frequencies $\Omega_n$ and $\omega_m$, respectively. As a result, in the interval of frequencies $\Omega\in(\Omega_n, \omega_n)$ a competition of the two resonances occurs and nontrivial effects can be found. Particularly, the evolution of contact angle has displayed the emergence of an additional local maximum at half a period, which is reminiscent of recent experimental observations,\cite{Noblin-Buguin-Brochard-Wyart-04} see Fig.~6(a).\\

\begin{acknowledgments}

We are grateful to S.~Shklyaev for many fruitful discussions, valuable comments, and providing the diagram shown in Fig.~\ref{fig4}.

I.F. is thankful to DAAD (Russian-German Mikhail Lomonosov Program, project No.~A/07/72463) for
support and to A.~Pikovsky for hosting the activity. A.S. was supported by German Science Foundation, DFG SPP
1164 ``Nano- and microfluidics,'' project 1021/1-2. The research has been a part of a joint German-Russian collaborative initiative recognized by German Science Foundation (DFG project No. 436 RUS113/977/0-1) and Russian
Foundation for Basic Research (RFBR project No.~08-01-91959). The authors gratefully acknowledge the funding organizations for support.

\end{acknowledgments}



\begin{thebibliography}{99}

\bibitem{deGennes-85}
P.-G.~De~Gennes,
\newblock ``Wetting: statics and dynamics,''
\newblock Rev. Mod. Phys. {\bf 57}, 827 (1985).

\bibitem{Leger-Joanny-92}
L.~Leger and J.~F.~Joanny,
\newblock ``Liquid spreading,''
\newblock Rep. Prog. Phys. {\bf 55}, 431 (1992).

\bibitem{Rauscher-Dietrich-08}
M.~Rauscher and S.~Dietrich
\newblock ``Wetting phenomena in nanofluidics,''
\newblock {Ann. Rev. Mater. Research} {\bf 38}, 143 (2008).


\bibitem{Lyubimov-Lyubimova-Shklyaev-06}
D.~V.~Lyubimov, T.~P.~Lyubimova, and S.~V.~Shklyaev,
\newblock ``Behavior of a drop on an oscillating solid plate,''
\newblock {Phys.~Fluids} {\bf 18}, 012101 (2006).

\bibitem{Vukasinovic-Smith-Glezer-07} B.~Vukasinovic, M.~K.~Smith, and~A.~Glezer,
``Dynamics of a sessile drop in forced vibration,''
J. Fluid Mech. {\bf 587}, 395 (2007).




\bibitem{Noblin-Buguin-Brochard-Wyart-04}
X.~Noblin, A.~Buguin, and F.~Brochard-Wyart,
\newblock ``Vibrated sessile drops: Transition between pinned and mobile contact line oscillations,''
\newblock {Eur. Phys. J. E} {\bf 14}, 395 (2004).

\bibitem{Brunet-Eggers-Deegan-07}
P.~Brunet, J.~Eggers, and R.~D.~Deegan,
\newblock ``Vibration-induced climbing of drops,''
\newblock Phys. Rev. Lett. {\bf 99}, 144501 (2007).

\bibitem{Daniel-Chaudhury-02}
S.~Daniel, S.~Sircar, J.~Gliem, and M.~K.~Chaudhury
\newblock ``Rectified motion of liquid drops on gradient surfaces induced by vibration,''
\newblock Langmuir {\bf 18}, 3404 (2002).

\bibitem{Daniel-etal-04}
S.~Daniel, S.~Sircar, J.~Gliem, and M.~K.~Chaudhury
\newblock ``Ratcheting motion of liquid drops on gradient surfaces,''
\newblock Langmuir {\bf 20}, 4085, (2004).

\bibitem{Mettu-Chaudhury-08}
S.~Mettu and M.~K.~Chaudhury,
\newblock ``Motion of Drops on a Surface Induced by Thermal Gradient and Vibration,''
\newblock {Langmuir} {\bf 24}, 10833 (2008).


\bibitem{Lyubimov-Lyubimova-Shklyaev-04}
D.~V.~Lyubimov, T.~P.~Lyubimova, and S.~V.~Shklyaev,
\newblock ``Non-axisymmetric oscillations of a hemispherical drop,''
\newblock {Fluid Dyn.} {\bf 39}, 851 (2004).

\bibitem{Alabuzhev-Lyubimov-07}
A.~A.~Alabuzhev and D.~V.~Lyubimov,
\newblock ``Effect of the contact-line dynamics on the natural oscillations of a cylindrical droplet,''
\newblock J. Appl. Mech. Tech. Phys. {\bf 48}, 686 (2007).

\bibitem{Shklyaev-Straube-08}
S.~V.~Shklyaev, A.~V.~Straube,
\newblock ``Linear oscillations of a compressible hemispherical bubble on a solid substrate,''
\newblock Phys.~Fluids {\bf 20}, 052102 (2008).


\bibitem{Daniel-Chaudhury-deGennes-05}
S.~Daniel, M.~K.~Chaudhury, and P.~G.~de~Gennes,
\newblock ``Vibration-actuated drop motion on surfaces for batch microfluidic processes,''
\newblock {Langmuir} {\bf 21}, 4240 (2005).

\bibitem{Buguin-Brochard-deGennes-06}
A.~Buguin,~F.~Brochard, and P.-G.~de~Gennes,
\newblock ``Motions induced by asymmetric vibrations. The solid/solid case,''
\newblock {Eur. Phys. J. E} {\bf 19}, 31 (2006).

\bibitem{Brochard-deGennes-07}
F.~Brochard-Wyart and P.-G.~de~Gennes,
\newblock ``Naive model for stick-slip processes,''
\newblock {Eur. Phys. J. E} {\bf 23}, 439 (2007).

\bibitem{Fleishman-Asscher-Urbakh-07}
D.~Fleishman,~Y.~Asscher, and M.~Urbakh,
\newblock ``Directed transport induced by asymmetric surface vibrations: making use of friction,''
\newblock {J. Phys.: Cond. Matt.} {\bf 19}, 096004 (2007).


\bibitem{Hocking-hyster-87}
L.~M.~Hocking,~
\newblock ``Waves produced by a vertically oscillating plate,''
\newblock {J.~Fluid Mech.}~{\bf 179},~267~(1987).

\bibitem{Hocking-87}
L.~M.~Hocking,
\newblock ``The damping of capillary-gravity waves at a rigid boundary,''
\newblock {J. Fluid Mech.} {\bf 179}, 253 (1987).




\bibitem{Mei-Liu-73}
C.~C.~Mei and L.~F.~Liu,
\newblock ``The damping of surface gravity waves in a bounded liquid,''
\newblock {J. Fluid Mech.} {\bf 59}, 239 (1973).

\bibitem{Henon-82}
M.~H\'enon
\newblock ``On the numerical computation of Poincar\'e maps,''
\newblock {Physica D}~{\bf 5},~412~(1982).












\end{thebibliography}
\end{document}